\documentclass[conference]{IEEEtran}
\IEEEoverridecommandlockouts
\usepackage[dvips]{graphicx}
\usepackage[fleqn]{amsmath}
\usepackage[varg]{txfonts}
\usepackage{colortbl}
\usepackage{algorithm}
\usepackage{algorithmic}
\usepackage{tabularx}
\usepackage{cite}
\usepackage{graphicx} 
\ifCLASSINFOpdf
\else
\fi
\begin{document}
%
\title{Distributed Interference-Aware Energy-Efficient Resource Allocation for Device-to-Device Communications Underlaying Cellular Networks}

\author{\IEEEauthorblockN{Zhenyu Zhou$^1$, Mianxiong Dong$^2$, Kaoru Ota$^3$, Jun Wu$^4$, Takuro Sato$^5$}
\IEEEauthorblockA{$^1$State Key Laboratory of Alternate Electrical Power System with Renewable Energy Sources,\\
School of Electrical and Electronic Engineering, North China Electric Power University, Beijing, China, 102206\\
$^2$National Institute of Information and Communications Technology, Kyoto, Japan\\
$^3$Department of Information and Electric Engineering, Muroran Institute of Technology, Muroran, Hokkaido, Japan\\
$^4$School of Information Security Engineering, Shanghai Jiao Tong University, Shanghai, China\\
$^5$Graduate School of Global Information and Telecommunication Studies, Waseda University, Tokyo, Japan}}

\maketitle



%
\IEEEpeerreviewmaketitle

\begin{abstract}
The introduction of device-to-device (D2D) into cellular networks poses many new challenges in the resource allocation design due to the co-channel interference caused by spectrum reuse and limited battery life of user equipments (UEs). In this paper, we propose a distributed interference-aware energy-efficient resource allocation algorithm to maximize each UE's energy efficiency (EE) subject to its specific quality of service (QoS) and maximum transmission power constraints. We model the resource allocation problem as a noncooperative game, in which each player is self-interested and wants to maximize its own EE. The formulated EE maximization problem is a non-convex problem and is transformed into a convex optimization problem by exploiting the properties of the nonlinear fractional programming. An iterative optimization algorithm is proposed and verified through computer simulations.
\end{abstract}

\section{Introduction}
Device-to-device (D2D) communications underlaying cellular networks bring numerous benefits including the proximity gain, the reuse gain, and the hop gain \cite{D2D_design}, increase the total throughput of the overall cellular network, \cite{D2D_LTE}, and can fit perfectly for future ubiquitous radio network \cite{Chorus_ZHOU}. However, the introduction of D2D communications into cellular networks poses many new challenges in the resource allocation design due to the co-channel interference caused by spectrum reuse and limited battery life of user equipments (UEs). Most of the previous studies mainly focus on how to maximize the spectral efficiency (SE) and ignore the energy consumption of UEs (see \cite{Zhaoyang_ICC2013, Feiran_WCNC2013, Doppler_TWC, Song_JSAC} and references therein). Only a limited amount of works have considered the energy efficiency (EE) optimization problem. In practical implementation, UEs are typically handheld devices with limited battery life and can quickly run out of battery if the energy consumption is ignored in the system design. Therefore, in this paper, we focus on how to optimize the EE through resource allocation in an interference-limited environment. 

For the EE optimization problem, distributed resource allocation algorithms which are based on either the reverse iterative combinatorial auction (ICA) game or the bisection method were proposed in \cite{Feiran_2012} and \cite{EE_analysis} respectively. However, the authors have not considered the QoS provisioning constraints and have not derived a close-form solution. Centralized resource allocation algorithms for optimizing the EE in the device-to-multidevice (D2MD) or D2D-cluster scenarios were proposed in \cite{D2D_MIMO} and \cite{Si_cluster} respectively. One major disadvantage of the centralized algorithms is that the computational complexity and signaling overhead increase significantly with the number of UEs.  Besides, since the optimization process is carried out in the BS, the optimal solution needs to be delivered to the UEs within the channel coherence time. Instead of maximizing EE, an auction-based resource allocation algorithm was proposed to maximize the battery lifetime in \cite{Feiran_ICC2013}, but cellular UEs were not taken into consideration. A coalition game based resource sharing algorithm was proposed in \cite{Wu_TVT2014} to jointly optimize the model selection and resource scheduling. The authors assumed that independent D2D UEs and cellular UEs can communicate with one another and act together as one entity to improve their EE in the game.

In this paper, we propose a distributed interference-aware energy-efficient resource allocation algorithm to maximize each UE's EE subject to the QoS provisioning and transmission power constraints. We model the resource allocation problem as a noncooperative game. Compared to the cooperative game model used in \cite{Wu_TVT2014}, the noncooperative model has the advantage of a lower overhead for information exchange between UEs. Both of the D2D UEs and cellular UEs are taken into consideration. The EE utility function of each player is defined as the SE divided by the total power consumption, which includes both transmission and circuit power. The formulated EE maximization problem is a non-convex problem and is transformed into a convex optimization problem by using the nonlinear fractional programming developed in \cite{Dinkelbach}. A Nash equilibrium is proved to exist in the noncooperative game. An iterative optimization algorithm is proposed to find the Nash equilibrium and is verified through computer simulations. EE and SE tradeoffs of the proposed algorithm are studied in \cite{ZhouWCL2014}

The structure of this paper is organized as follows: Section \ref{System Model} introduces the system model of the D2D communication underlaying cellular networks. Section \ref{central} introduces the centralized resource allocation scenario and provides some insights by considering several special cases. Section \ref{distributed} introduces the distributed iterative optimization algorithm for maximizing each UE's EE. Section \ref{Simulation Results} introduces the simulation parameters, results and analyses. Section \ref{Conclusion} gives the conclusion.

\section{System Model}
\label{System Model}

\begin{figure}[t]
\begin{center}
\scalebox{0.36} 
{\includegraphics{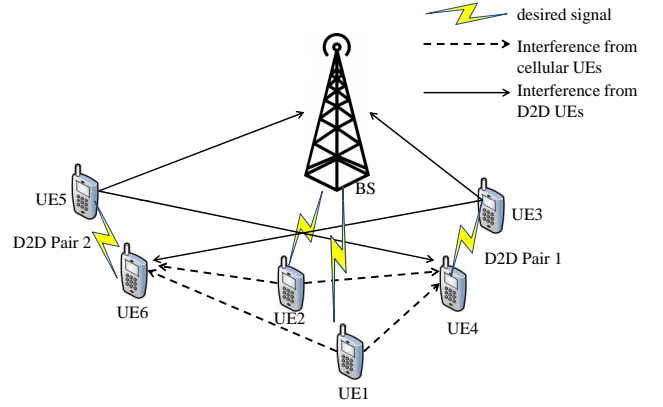}}
\end{center}
\caption{System model of D2D communications with uplink channel reuse.}
\label{D2D_model}
\end{figure}

In this paper, we consider the uplink scenario of a single cellular network, which is composed of the base station, the D2D UEs, and the cellular UEs. Fig. \ref{D2D_model} shows the system model of the D2D communications with uplink resource sharing. There are two cellular UEs ($\mbox{UE}_1$ and $\mbox{UE}_2$), and two D2D pairs ($\mbox{UE}_3$ and $\mbox{UE}_4$, and $\mbox{UE}_5$ and $\mbox{UE}_6$ respectively). A pair of D2D transmitter and receiver form a D2D link, and a cellular UE and the BS form a cellular link. The UEs in a D2D pair are close enough to enable D2D communication. Each cellular UE is allocated with an orthogonal link (e.g., an orthogonal resource block in LTE), i.e., there is no co-channel interference between cellular UEs. At the same time, the two D2D pairs reuse the same channels allocated to cellular UEs in order to improve the spectrum efficiency. As a result, the BS suffers from the interference caused by the D2D transmitters ($\mbox{UE}_3$ and $\mbox{UE}_5$), and the D2D receiver ($\mbox{UE}_4$ and $\mbox{UE}_6$) suffers from the interference caused by cellular UEs ($\mbox{UE}_1$ and $\mbox{UE}_2$) and the other D2D transmitters that reuse the same channel ($\mbox{UE}_5$ or $\mbox{UE}_3$ respectively).

The set of UEs is denoted as $\mathcal{S}=\{ \mathcal{N}, \mathcal{K} \}$, where $\mathcal{N}$ and $\mathcal{K}$ denote the sets of D2D UEs and cellular UEs respectively. The total number of D2D links and cellular links are denoted as $N$ and $K$ respectively. The signal to interference plus noise ratio (SINR) of the $i$-th D2D pair ($i \in \mathcal{N}$) in the $k$-th ($k \in \mathcal{K}$) channel is given by
\begin{align}
\label{eq:SINRD}
\gamma_i^k = \frac{p_i^k g_{i}^k}{p_c^k g_{c, i}^k+\sum_{j=1, j\neq i}^{N}p_{j}^k g_{j, i}^k+N_0}, 
\end{align}
where $p_i^k$, $p_c^k$, and $p_{j}^k$ are the transmission power of the $i$-th D2D transmitter, the $k$-th cellular UE, and the $j$-th D2D transmitter in the $k$-th channel respectively. $g_{i}^k$ is the channel gain of the $i$-th D2D pair, $g^k_{c, i}$ is the interference channel gain between the $k$-th cellular UE and the $i$-th D2D receiver, and $g_{j, i}^k$ is the interference channel gain between the $j$-th D2D transmitter and the $i$-th D2D receiver. $N_0$ is the nosier power. $p_c^k g^k_{c, i}$ and $\sum_{j=1, j\neq i}^{N} p_{j}^k g_{j, i}^k$ denote the interference from the cellular UE and the other D2D pairs that reuse the $k$-th channel respectively.  

The received SINR of the $k$-th cellular UE at the BS is given by
\begin{align}
\label{eq:SINRC}
\gamma_c^k = \frac{p_c^k g_c^k}{\sum_{i=1}^{N}p_{i}^k g_{i, c}^k+N_0}, 
\end{align}
where $g_c^k$ is the channel gain between the $k$-th cellular UE and the BS, $g^k_{i, c}$ is the interference channel gain between the $i$-th D2D transmitter and the BS in the $k$-th channel. $\sum_{i=1}^{N}p_{i}^k g_{i, c}^k$ denote the interference from all of the D2D pairs to the BS in the $k$-th channel.  

The achievable rates of the $i$-th D2D pair and the $k$-th cellular UE are given by
\begin{align}
\label{eq:rateD}
r_i^d &=\sum_{k=1}^K \log_2 \left( 1+\gamma_i^k \right),\\
\label{eq:rateC}
r_k^c&=\log_2 \left( 1+\gamma_c^k \right).
\end{align}
 The total power consumption of the $i$-th D2D pair and the $k$-th cellular UE are given by
 \begin{align}
 \label{eq:powerD}
 p_{i,total}^d &= \sum_{k=1}^K \frac{1}{\eta } p_i^k+2p_{cir},\\
 \label{eq:powerC}
  p_{k, total}^c &= \frac{1}{\eta} p_c^k+p_{cir},
 \end{align}
 where $p_{i, total}^d$ is the total power consumption of the $i$-th D2D pair, which is composed of the transmission power over all of the $K$ channels, i.e., $\sum_{k=1}^K \frac{1}{\eta } p_i^k$, and the circuit power of both the D2D transmitter and receiver, i.e., $2p_{cir}$. The circuit power of any UE is assumed as the same and denoted as $p_{cir}$. $\eta$ is the Power Amplifier (PA) efficiency, i.e., $0 < \eta < 1$. $p_{k, total}^c$ is the total power consumption of the $k$-th cellular UE, which is composed of the transmission power $\frac{1}{\eta} p_c^k$ and the circuit power only at the transmitter side. The power consumption of the BS is not taken into consideration. 

 \section{Centralized Interference-Aware Energy-Efficient Resource Allocation}
\label{central}
The EE introduced in \cite{EE_definition} is defined as the SE divided by the total power consumed, i.e., bit/Hz/J. In this section, we study the centralized energy-efficient resource allocation method, which is employed at the BS. The EE of the overall network is a function of the power allocation strategies, which is given by
\begin{align}
 \label{eq:U_EE}
 U_{EE}(\mathcal{P}_d, \mathcal{P}_c)=\sum_{i=1}^N \frac{r_i^d}{ p_{i,total}^d }+\sum_{k=1}^K \frac{r_k^c}{p_{k, total}^c}, 
\end{align}
where $\mathcal{P}_d$ and $\mathcal{P}_c$ are the sets of power allocation strategies for the D2D UEs and cellular UEs respectively, i.e., $\mathcal{P}_d=\{p_i^k \mid 0 \leq  \sum_{k=1}^K p_i^k  \leq  p_{i, max}^d,    i \in \mathcal{N}, k \in \mathcal{K}\}$, $\mathcal{P}_c=\{p_k^c \mid 0 \leq p_c^k \leq p_{k, max}^c, k \in \mathcal{K}\}$. $ p_{i, max}^d$ and $ p_{k, max}^c$ are the maximum transmission power of the $i$-th D2D transmitter and the $k$-th cellular UE respectively. 
This definition of (\ref{eq:U_EE}) is not based on the ratio of sum network throughput to sum network power consumption as in \cite{Feiran_2012, Wu_TVT2014} because transmission power and achievable rates can not be shared among UEs \cite{Miao_TWC2011}. Taking (\ref{eq:SINRD}), (\ref{eq:SINRC}), (\ref{eq:rateD}), (\ref{eq:rateC}), (\ref{eq:powerD}), and (\ref{eq:powerC}) into (\ref{eq:U_EE}), the EE of the overall network is rewritten as
\begin{align}
 \label{eq:U_EE_full}
 U_{EE}(\mathcal{P}_d, \mathcal{P}_c) &=\sum_{i=1}^N \frac{\sum_{k=1}^K \log_2 \left( 1+\frac{p_i^k g_{i}^k}{p_c^k g^k_{c, i}+\sum_{j=1, j\neq i}^{N}p_{j}^k g_{j, i}^k+N_0} \right) }{\sum_{k=1}^K \frac{1}{\eta } p_i^k+2p_{cir}}\notag\\
 &+\sum_{k=1}^K \frac{\log_2 \left( 1+\frac{p_c^k g_c^k}{\sum_{i=1}^{N}p_{i}^k g_{i, c}^k+N_0} \right)}{\frac{1}{\eta} p_c^k+p_{cir}}.
\end{align}
The $U_{EE}$ defined in (\ref{eq:U_EE_full}) is not a concave function for $p_i^k, p_c^k$ ($p_i^k \in \mathcal{P}_d, p_c^k \in \mathcal{P}_c)$, and it is intractable to find the global maximum EE of the overall network. However, we can get some insights about energy-efficient power allocation design by considering some special cases.
 
\subsection{The Circuit Power Dominated Case}
\label{circuit_case}

The circuit power dominated case represents that $p_{cir}>>p_i^k, p_c^k$, $\forall i \in \mathcal{N}, \forall k \in \mathcal{K}$. The circuit power dominated case arises when the transmitter is close to the receiver and little transmission power is needed to satisfy the QoS requirement. The $U_{EE}$ defined in (\ref{eq:U_EE_full}) is rewritten as
\begin{align}
 &U_{EE}(\mathcal{P}_d, \mathcal{P}_c) \approx \notag\\
&\frac{1}{2p_{cir}} \bigg[\sum_{i=1}^N \sum_{k=1}^K \log_2 \bigg( 1+\frac{p_i^k g_{i}^k}{p_c^k g^k_{c, i}+\sum_{j=1, j\neq i}^{N}p_{j}^k g_{j, i}^k+N_0} \bigg) \notag\\
 &+\sum_{k=1}^K 2\log_2 \bigg( 1+\frac{p_c^k g_c^k}{\sum_{i=1}^{N}p_{i}^k g_{i, c}^k+N_0} \bigg) \bigg].
\end{align}
The EE maximization problem in the circuit power dominated case is equivalent to the conventional sum rate maximization problem, which has been discussed in \cite{Zhaoyang_ICC2013, Feiran_WCNC2013, Doppler_TWC, Song_JSAC}. 

\subsection{The Transmission Power Dominated Case}
\label{tran_case}
The transmission power dominated case represents that $p_i^k, p_c^k >> p_{cir}$, $\forall i \in \mathcal{N}, \forall k \in \mathcal{K}$. This case arises in long-range communication and interference-limited scenarios where large transmission power is required to compensate for the degradation of the received SINR. The $U_{EE}$ defined in (\ref{eq:U_EE_full}) can be rewritten as
\begin{align}
 U_{EE} (\mathcal{P}_d, \mathcal{P}_c) & \approx \sum_{i=1}^N \frac{\sum_{k=1}^K \log_2 \left( 1+\frac{p_i^k g_{i}^k}{p_c^k g^k_{c, i}+\sum_{j=1, j\neq i}^{N} p_{j}^k g_{j, i}^k+N_0} \right) }{\sum_{k=1}^K \frac{1}{\eta } p_i^k}\notag\\
 &+\sum_{k=1}^K \frac{\log_2 \left( 1+\frac{p_c^k g_c^k}{\sum_{i=1}^{N}p_{i}^k g_{i, c}^k+N_0} \right)}{\frac{1}{\eta} p_c^k}.
\end{align}
Since $U_{EE}$ is strictly decreasing with $p_i^k, p_c^k$ (it can be proved that $\frac{\partial U_{EE}}{p_i^k}<0$ and $\frac{\partial U_{EE}}{p_c^k}<0$, $\forall i,k$), the optimal strategy is to use as little power as possible subject to the QoS constraint. 

\subsection{Noise Dominated Case}
\label{noise_case}

The noise dominated case represents that $N_0>>p_c^k g_{c, i}^k+\sum_{j=1, j\neq i}^N p_j^k g_{j, i}^k$, $N_0>>\sum_{i=1}^N p_i^k g_{i,c}^k$, $\forall i \in \mathcal{N}, \forall k \in \mathcal{K}$. The $U_{EE}$ defined in (\ref{eq:U_EE_full}) can be rewritten as
\begin{align}
 U_{EE}(\mathcal{P}_d, \mathcal{P}_c) \approx \!\!\sum_{i=1}^N \! \frac{\sum_{k=1}^K \log_2 \left( 1+\frac{p_i^k g_{i}^k}{N_0} \right) }{\sum_{k=1}^K \frac{1}{\eta } p_i^k+2p_{cir}} +\sum_{k=1}^K \! \frac{\log_2 \left( 1+\frac{p_c^k g_c^k}{N_0} \right)}{\frac{1}{\eta} p_c^k+p_{cir}}.
\end{align}
Thus, the EE maximization problem in the noise dominated case is decomposed into independent $N+K$ subproblems, in which each UE tries to maximize its own EE without considering the other UEs' strategies.

\subsection{Interference Dominated Case}
\label{interference_case}
 
In the interference dominated case, the interference is much stronger than the noise, i.e., $p_c^k g_{c, i}^k+\sum_{j=1, j\neq i}^N p_j^k g_{j, i}^k>>N_0$, $\sum_{i=1}^N p_i^k g_{i,c}^k>>N_0$, $\forall i \in \mathcal{N}, \forall k \in \mathcal{K}$. The $U_{EE}$ defined in (\ref{eq:U_EE_full}) can be rewritten as
\begin{align}
 U_{EE}(\mathcal{P}_d, \mathcal{P}_c)& \approx \sum_{i=1}^N \frac{\sum_{k=1}^K \log_2 \left( 1+\frac{p_i^k g_{i}^k}{p_c^k g_{c, i}^k+\sum_{j=1, j\neq i}^N p_j^k g_{j, i}^k} \right) }{\sum_{k=1}^K \frac{1}{\eta } p_i^k+2p_{cir}}\notag\\
 &+\sum_{k=1}^K \frac{\log_2 \left( 1+\frac{p_k^c g_k^c}{\sum_{i=1}^N p_i^k g_{i,c}^k} \right)}{\frac{1}{\eta} p_c^k+p_{cir}}.
\end{align}
The EE in the interference dominated case is maximized by only allowing the UE (either the cellular UE or the D2D UE) with the highest channel gain to transmit since that $\log_2 \left( 1+\frac{p_i^k g_{i}^k}{p_c^k g_{c, i}^k+\sum_{j=1, j\neq i}^N p_j^k g_{j, i}^k} \right) \to \infty$ as $p_c^k \to 0$, $p_j^k \to 0$.

\subsection{Cellular UE Dominated Case}
\label{C_case}

The cellular UE dominated case arises in scenarios where a cellular UE is far from the BS but close to the D2D pair, and the transmission power of cellular UEs is much stronger than the transmission power of the D2D transmitter, i.e., $p_c^k>>p_i^k$, $\forall i \in \mathcal{N}, \forall k \in \mathcal{K}$. The $U_{EE}$ defined in (\ref{eq:U_EE_full}) can be rewritten as
\begin{align}
 U_{EE}(\mathcal{P}_d, \mathcal{P}_c)& \approx \sum_{k=1}^K \frac{\log_2 \left( 1+\frac{p_c^k g_c^k}{N_0} \right)}{\frac{1}{\eta} p_c^k+p_{cir}}.
\end{align}
The D2D UEs are forced to stop transmission due to the severe interference caused by cellular UEs, which solely occupy all of the available channels. The optimum solution can be obtained by using the bisection method \cite{convex_optimization}.

\subsection{D2D UEs Dominated Case}
\label{D_case}
In the D2D UEs dominated case, we have $p_i^k>>p_c^k$, $\forall i \in \mathcal{N}, \forall k \in \mathcal{K}$. The $U_{EE}$ defined in (\ref{eq:U_EE_full}) can be rewritten as
\begin{align}
 U_{EE}(\mathcal{P}_d, \mathcal{P}_c)& \approx \sum_{i=1}^N \frac{\sum_{k=1}^K \log_2 \left( 1+\frac{p_i^k g_{i}^k}{\sum_{j=1, j\neq i}^{N} p_{j}^k g_{j, i}^k+N_0} \right) }{\sum_{k=1}^K \frac{1}{\eta } p_i^k+2p_{cir}}.
\end{align}
The cellular UEs are forced to stop transmission due to the severe interference caused by D2D UEs, which solely occupy all of the available channels. The optimum solution can be obtained by using the bisection method \cite{convex_optimization}.

\section{Distributed Interference-Aware Energy-Efficient Resource Allocation}
\label{distributed}

\subsection{Problem Formulation}

In the centralized resource allocation, the BS requires the complete network knowledge and the computational complexity and signaling overhead increase significantly with the number of UEs. Therefore, in this section, we focus on the more practical distributed resource allocation problem, which is modeled as a noncooperative game. 

In the noncooperative game, each UE is self-interested and wants to maximize its own EE. The strategy set of the $i$-th D2D transmitter is denoted as $\mathbf{p}_i^d=\{p_i^k \mid 0 \leq  \sum_{k=1}^K p_i^k  \leq  p_{i, max}^d, k \in \mathcal{K} \}$, $\forall i \in \mathcal{N}$. The strategy set of the $k$-th 
cellular UE is denoted as $\mathbf{p}_k^c=\{ p_c^k \mid 0 \leq p_c^k \leq p_{k, max}^c  \}$, $\forall k \in \mathcal{K}$. The strategy set of the other D2D transmitters in $\mathcal{N} \backslash  \{i\}$ is denoted as $\mathbf{p}_{-i}^d=\{ p_j^k \mid  0 \leq  \sum_{k=1}^K p_j^k  \leq  p_{j, max}^d, k \in \mathcal{K}, j \in \mathcal{N}, j \neq i \}$, $\forall i \in \mathcal{N}$. The strategy set of the other cellular UEs in $\mathcal{K} \backslash \{k\}$ is denoted as $\mathbf{p}_{-k}^c=\{p_c^m \mid 0 \leq p_c^m \leq p_{m, max}^c, m\in \mathcal{K}, m\neq k  \}$, $\forall k \in \mathcal{K}$.


For the $i$-th D2D pair, its EE $U_i^d$ depends not only on $\mathbf{p}_i^d$, but also on the strategies taken by other UEs in $\mathcal{S}\backslash \{i\}$, i.e.,  $\mathbf{p}_{-i}^d, \mathbf{p}_k^c, \mathbf{p}_{-k}^c$. $U_i^d$ is defined as
\begin{align}
\label{eq:UE_EED}
&U_i^d (\mathbf{p}_i^d, \mathbf{p}_{-i}^d, \mathbf{p}_k^c, \mathbf{p}_{-k}^c)\notag\\
&=\frac{r_i^d}{p_{i, total}^d}=\frac{\sum_{k=1}^K \log_2 \left( 1+\frac{p_i^k g_{i}^k}{p_c^k g^k_{c, i}+\sum_{j=1, j\neq i}^{N}p_{j}^k g_{j, i}^k+N_0} \right) }{\sum_{k=1}^K \frac{1}{\eta } p_i^k+2p_{cir}}.
\end{align}
Therefore, the EE maximization problem of the $i$-th D2D pair is formulated as
\begin{align}
\label{eq:Dproblem}
&\max. \hspace{10mm} U_i^d (\mathbf{p}_i^d, \mathbf{p}_{-i}^d, \mathbf{p}_k^c, \mathbf{p}_{-k}^c) \notag\\
&\mbox{s.t.} \hspace{15mm} C1, C2.
\end{align}
\begin{align}
C1: &r_i^d \geq R_{i, min}^d, \\
C2: &0 \leq \sum_{k=1}^K p_i^k \leq p_{i, max}^d.
\end{align}
Similarly, the EE of the $k$-th cellular UE $U_k^c$ is defined as
\begin{align}
U_k^c (\mathbf{p}_i^d, \mathbf{p}_{-i}^d, \mathbf{p}_k^c, \mathbf{p}_{-k}^c)=\frac{r_k^c}{p_{k, total}^c}=\frac{\log_2 \left( 1+\frac{p_c^k g_c^k}{\sum_{i=1}^{N}p_{i}^k g_{i, c}^k+N_0} \right)}{\frac{1}{\eta} p_c^k+p_{cir}}.
\end{align}
The corresponding EE maximization problem is formulated as
\begin{align}
\label{eq:Cproblem}
&\max. \hspace{10mm} U_k^c (\mathbf{p}_i^d, \mathbf{p}_{-i}^d, \mathbf{p}_k^c, \mathbf{p}_{-k}^c) \notag\\
&\mbox{s.t.} \hspace{15mm} C3, C4.
\end{align}
\begin{align}
C3: &r_k^c \geq R_{k, min}^c,\\
C4: &0 \leq p_c^k \leq p_{k, max}^c.
\end{align}
The constraint C1 and C3 specify the QoS requirements in terms of minimum transmission rate. C2 and C4 are the non-negative constraints on the power allocation variables.

\subsection{The Objective Function Transformation}
\label{transformation}

The objective functions in (\ref{eq:Dproblem}) and (\ref{eq:Cproblem}) are non-convex due to the fractional form. In order to derive a closed-form solution, we transformed the fractional objective function to a convex optimization function by using the nonlinear fractional programming developed in \cite{Dinkelbach}. We define the maximum EE of the $i$-th D2D pair as $q^{d*}_i$, which is given by
\begin{equation}
q^{d*}_i=\max. U_i^d (\mathbf{p}_i^d, \mathbf{p}_{-i}^d, \mathbf{p}_k^c, \mathbf{p}_{-k}^c)=\frac{r_i^d(\mathbf{p}_i^{d*})}{p_{i, total}^d(\mathbf{p}_i^{d*})}.
\end{equation} 
where $\mathbf{p}_i^{d*}$ is the best response of the $i$-th D2D transmitter given the other UEs' strategies $\mathbf{p}_{-i}^d, \mathbf{p}_k^c, \mathbf{p}_{-k}^c$. The following theorem can be proved:

\textbf{\emph{Theorem 1:}} The maximum EE $q_i^{d*}$ is achieved if and only if 
\begin{align}
\max. \:\: r_i^d (\mathbf{p}_i^d)-q_i^{d*}p_{i, total}^d(\mathbf{p}_i^d)=r_i^d (\mathbf{p}_i^{d*})-q_i^{d*}p_{i, total}^d(\mathbf{p}_i^{d*})=0.
\end{align}

\emph{Proof:} The proof of Theorem 1 is similar to the proof of the Theorem (page 494 in \cite{Dinkelbach}).

Similarly, for the maximum EE of the $k$-th cellular UE $q^{c*}_k$, we will have similar theorem as \textbf{\emph{Theorem 1:}}

\textbf{\emph{Theorem 2:}} The maximum EE $q_k^{c*}$ is achieved if and only if 
\begin{align}
\max. \:\: r_k^c (\mathbf{p}_k^c)-q_k^{c*}p_{k, total}^c(\mathbf{p}_k^c)=r_k^c (\mathbf{p}_k^{c*})-q_k^{c*}p_{k, total}^c(\mathbf{p}_k^{c*})=0.
\end{align}
$\mathbf{p}_k^{c*}$ is the best response of the $k$-th cellular UE given the other UEs' strategies $\mathbf{p}_i^d, \mathbf{p}_{-i}^d, \mathbf{p}_{-k}^c$. 

\subsection{The Iterative Optimization Algorithm}
\label{algorithm} 
The proposed algorithm is summarized in Algorithm \ref{offline algorithm}. $n$ is the iteration index, $L_{max}$ is the maximum number of iterations, and $\Delta$ is the maximum tolerance. 
At each iteration, for any given $q_i^{d}$ or $q_k^{c}$, the resource allocation strategy for the D2D UE or the cellular UE can be obtained by solving the following transformed optimization problems respectively:
\begin{align}
 \label{eq:transformed problemD}
  &\max . \:\: r_i^d (\mathbf{p}_i^d)-q_i^d p_{i, total}^d(\mathbf{p}_i^d) \nonumber\\
 &\mbox{s.t.} \:\:\: C1, C2.
 \end{align}
 \begin{align}
 \label{eq:transformed problemC}
  &\max . \:\: r_k^c (\mathbf{p}_k^c)-q_k^c p_{k, total}^c (\mathbf{p}_k^c) \nonumber\\
 &\mbox{s.t.} \:\:\: C3, C4.
 \end{align}

\begin{algorithm}[t]
\caption{Iterative Resource Allocation Algorithm}
\label{offline algorithm}
\begin{algorithmic}[1]
\STATE $q_i^d \leftarrow 0$, $q_k^c \leftarrow 0$, $L_{max} \leftarrow 10$, $n \leftarrow 1$, $\Delta \leftarrow 10^{-3}$ 
\FOR{$n=1$ to $L_{max}$}
\IF {D2D link}
\STATE solve (\ref{eq:transformed problemD}) for a given $q_i^d$ and obtain the set of strategies $\mathbf{p}_i^d$
\IF{$r_i^d(\mathbf{p}_i^d)-q_i^d p_{i,total}^d (\mathbf{p}_i^d) \leq \Delta$,}
 \STATE $\mathbf{p}_i^{d*}=\mathbf{p}_i^d$, and $\displaystyle q_i^{d*}=\frac{r_i^d(\mathbf{p}_i^{d*})}{p_{i, total}^d(\mathbf{p}_i^{d*})}$ 
\STATE \textbf{break}
\ELSE
\STATE $\displaystyle q_i^d=\frac{r_i^d(\mathbf{p}_i^d)}{p_{i, total}^d(\mathbf{p}_i^d)}$, and $n=n+1$
\ENDIF
\ELSE
\STATE solve (\ref{eq:transformed problemC}) for a given $q_k^c$ and obtain the set of strategies $\mathbf{p}_k^c$
\IF{$r_k^c(\mathbf{p}_k^c)-q_k^c p_{k,total}^c (\mathbf{p}_k^c) \leq \Delta$,}
 \STATE $\mathbf{p}_k^{c*}=\mathbf{p}_c$, and $\displaystyle q_k^{c*}=\frac{r_k^c (\mathbf{p}_k^{c*})}{p_{k, total}^c(\mathbf{p}_k^{c*})}$ 
\STATE \textbf{break}
\ELSE
\STATE $\displaystyle q_k^c=\frac{r_k^c(\mathbf{p}_k^c)}{p_{k, total}^c(\mathbf{p}_k^c)}$, and $n=n+1$
\ENDIF
\ENDIF
\ENDFOR
\end{algorithmic}
\end{algorithm}

Taking the D2D UEs as an example, the Lagrangian associated with the problem (\ref{eq:transformed problemD}) is given by
   \begin{align}
\mathcal{L}_{EE}(\mathbf{p}_i^d, \alpha_i, \beta_i) &=r_i^d (\mathbf{p}_i^d)-q_i^d p_{i, total}^d (\mathbf{p}_i^d)\notag\\
&+\alpha_i \left( r_i^d-R_{i, min}^d \right)-\beta_i \left( \sum_{k=1}^K p_i^k-p_{i, max}^d\right),
 \end{align}
 where $\alpha_i$, $\beta_i$ are the Lagrange multipliers associated with the constraints C1 and C2 respectively. The constraint $p_i^k \geq 0$ is absorbed into the Karush-Kuhn-Tucker (KKT) condition when solving the equivalent Lagrange dual problem:
 \begin{equation}
\label{eq:dual problem}
 \displaystyle \min_{\displaystyle (\alpha_i \geq 0, \beta_i \geq 0)}\!\!\!\!. \hspace{5mm} \max_{\displaystyle (\mathbf{p}_i^d)}. \:\:\: \mathcal{L}_{EE}(\mathbf{p}_i^d, \alpha_i, \beta_i) 
\end{equation}
It is noted that the objective function in (\ref{eq:transformed problemD}) is a concave function of $\mathbf{p}_i^d$, and the primal and dual optimal points forms an saddle-point of the Lagrangian. The dual problem in (\ref{eq:dual problem}) can be decomposed into two subproblems: the maximization problem solves the power allocation problem to find the best strategy and the minimization problem solves the master dual problem to find the corresponding Lagrange multiplier. For any given $q_i^d$, the solution is given by
\begin{equation}
\label{eq:waterfilling}
p_i^{k}=\left[ \frac{\eta(1+\alpha_i) \log_2e }{q_i^d+\eta\beta_i }-\frac{p_c^k g^k_{c, i}+\sum_{j=1, j\neq i}^{N}p_{j}^k g_{j, i}^k+N_0}{g_{i}^k}\right]^{+},
\end{equation}
where $[x]^+=\max\{0,x\}$. Equation (\ref{eq:waterfilling}) indicates a water-filling algorithm for transmission power allocation, and the interference from the other UEs decreases the water level. For solving the minimization problem, the Lagrange multipliers can be updated by using the gradient method \cite{improved_step_size, subgradient} as
\begin{align}
\alpha_i (\tau +1)&=\left[ \alpha_i(\tau )-\mu_{i, \alpha} (\tau ) \left( r_i^d(\tau )-R_{i, min}^d \right)    \right]^{+},\\
\beta_i (\tau +1)&=\left[ \beta_i(\tau )+\mu_{i, \beta} (\tau ) \left( \sum_{k=1}^K p_i^k (\tau) - p_{i, max}^d \right)    \right]^{+},
\end{align}
where $\tau$ is the iteration index, $\mu_{i, \alpha}, \mu_{i, \beta}$ are the positive step sizes. Similarly, the optimum solution of $p^c$ is given by
\begin{equation}
\label{eq:waterfilling_CE}
p_c^k=\left[ \frac{\eta (1-\delta_k) \log_2e }{q_k^c+\eta \theta_k  }-\frac{\sum_{i=1}^N p_{i}^k g_{i, c}^k+N_0}{g_{c}^k} \right]^+,
\end{equation}
where $\delta_k, \theta_k$ are the Lagrange multipliers associated with the constraints C3 and C4 respectively.

 A Nash equilibrium is a set of power allocation strategies that none UE (neither D2D UE nor cellular UE) can unilaterally improve its EE by choosing a different power allocation strategy, i.e., $\forall i \in \mathcal{N}, \forall k \in \mathcal{K}$, 
\begin{align}
 U_i^d(\mathbf{p}_i^{d*}, \mathbf{p}_{-i}^{d*}, \mathbf{p}^{c*}_k, \mathbf{p}_{-k}^{c*}) & \geq  U_i^d(\mathbf{p}_i^d, \mathbf{p}_{-i}^d, \mathbf{p}_k^c, \mathbf{p}_{-k}^c), \\
 U_k^c(\mathbf{p}_i^{d*}, \mathbf{p}_{-i}^{d*}, \mathbf{p}^{c*}_k, \mathbf{p}_{-k}^{c*})  & \geq  U_k^c(\mathbf{p}_i^{d}, \mathbf{p}_{-i}^d, \mathbf{p}^c_k, \mathbf{p}_{-k}^c).
 \end{align}

  \textbf{\emph{Theorem 3:}} 
 A Nash equilibrium exists in the noncooperaive game. Furthermore, the strategy set $\{ \mathbf{p}_i^{d*}, \mathbf{p}^{c*}_k \mid  i \in \mathcal{N},  k \in \mathcal{K}\}$ obtained by using Algorithm \ref{offline algorithm} is the Nash equilibrium.
 
\begin{IEEEproof}
The proof of Theorem 3 is given in \cite{LongWCL}.
\end{IEEEproof}

\section{Simulation Results}
\label{Simulation Results}

\begin{table}[t]
\caption{Simulation Parameters.}
\label{simulation_parameters}
\begin{center}
\begin{tabular}{|l|l|}
\hline
\textbf{Parameter}&\textbf{Value}\\
\hline
Cell radius & 500 m\\
\hline
Maximum D2D transmission distance & 25 m\\
\hline
Maximum transmission power $\displaystyle p_{i,max}^d, \displaystyle p_{k, max}^c$ & 200 mW (23 dBm)\\
\hline
Constant circuit power $p_{cir}$ &10 mW (10 dBm)\\
\hline
Thermal noise power $N_0$ & $\displaystyle 10^{-7}$ W\\
\hline
Number of D2D pairs $N$ & 5\\
\hline
Number of cellular UEs $K$ & 3\\
\hline
PA efficiency $\eta $ & 35\%\\
\hline
QoS of cellular UEs $\displaystyle R_{k, min}^c$ & 0.1 bit/s/Hz\\
\hline
QoS of D2D UEs $\displaystyle R_{i, min}^d$ & 0.5 bit/s/Hz\\
\hline
\end{tabular}
\end{center}
\end{table}

In this section, the proposed algorithm is verified through computer simulations. The values of simulation parameters are inspired by \cite{Feiran_WCNC2013, Feiran_2012, Song_JSAC} , and are summarized in Table \ref{simulation_parameters}. We compare the proposed EE maximization algorithm (labeled as ``energy-efficient") with the SE maximization algorithm (labeled as ``spectral-efficient" ) and the random power allocation algorithm (labeled as ``random"), whose details are given in \cite{ZhouIET2015}. The results are averaged through a total number of $1000$ simulations and  normalized by the maximum EE value of D2D links. For each simulation, the locations of the cellular UEs and D2D UEs are generated randomly within a cell with a radius of $500$ m. The maximum distance between any two D2D UEs that form a D2D pair is $25$ m. The channel gain between the transmitter $i$ and the receiver $j$ is calculated as $d_{i, j}^{-2} \mid h_{i, j} \mid^2$, where $d_{i, j}$ is the distance between the transmitter $i$ and the receiver $j$, $h_{i, j}$ is the complex Gaussian channel coefficient that satisfies $h_{i, j} \sim \mathcal{CN} (0, 1)$.

\begin{figure}[t]
\begin{center}
\scalebox{0.55} 
{\includegraphics{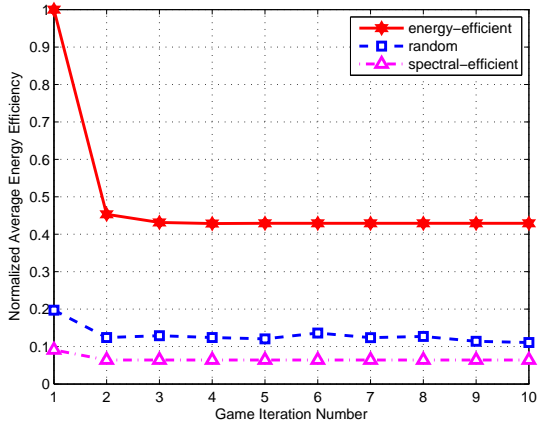}}
\end{center}
\caption{The normalized average energy efficiency of D2D links corresponding to the number of game iterations.}
\label{EE_D2D}
\end{figure}

\begin{figure}[t]
\begin{center}
\scalebox{0.55} 
{\includegraphics{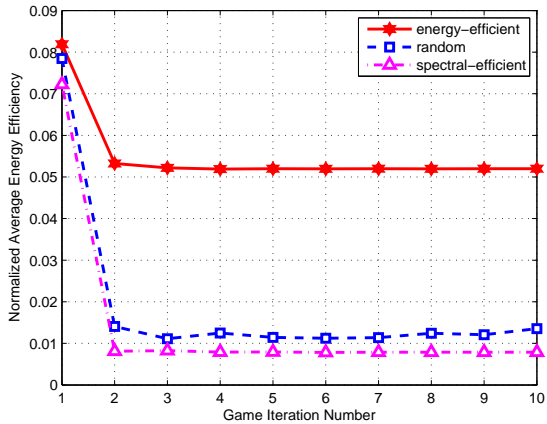}}
\end{center}
\caption{The normalized average energy efficiency of cellular links corresponding to the number of game iterations.}
\label{EE_CE}
\end{figure}

Fig. \ref{EE_D2D} shows the normalized average EE of D2D links corresponding to the number of game iterations. It is clear that the proposed energy-efficient algorithm significantly outperforms the conventional spectral-efficient algorithm and the random algorithm in terms of EE. The spectral-efficient algorithm has the worst EE performance among the three because power consumption is completely ignored in the optimization process. 

Fig. \ref{EE_CE} shows the normalized average EE of cellular links corresponding to the number of game iterations. The simulation results demonstrate that the proposed algorithm achieves the best performance again. Comparing Fig. \ref{EE_CE} with Fig. \ref{EE_D2D}, we find that the D2D links can achieve a much better EE than the cellular links due to the proximity gain and the channel  reuse gain. The proposed EE algorithm and the conventional SE algorithm converges to the equilibrium within $3 \sim 4$ game iterations, while the random algorithm fluctuates around the equilibrium since that the transmission power strategy is randomly selected. In the beginning of the game, we assume that channels are only used by cellular links. Hence, the EE of cellular links is highest in the first iteration due to the lack of interference. Then, D2D UEs enter into the game, and decides its optimum transmission power. The EE of D2D links is also highest in the first iteration since that the interference from cellular UEs and other D2D links is lowest in the first iteration.

\section{Conclusion}
\label{Conclusion}
In this paper, a distributed interference-aware energy-efficient resource allocation algorithm was proposed for D2D communications with uplink channel reuse. The close-form optimal solution was derived and was proved to be a Nash equilibrium. Simulation results verified that the proposed algorithm significantly outperforms conventional algorithms in terms of energy efficiency. 
 
\section*{Acknowledgment}

This work was partially supported by Fundamental Research Funds for the Central Universities under Grant Number 14MS08, NSFC Grant No. 61450110085, the Open Research Project of the State Key Laboratory of Industrial Control Technology, Zhejiang University, China (No. ICT1407), JSPS KAKENHI Grant Number 25880002, 26730056 and JSPS A3 Foresight Program, China Mobile Communication Co. Ltd. Research Institute (CMRI), and China Electric Power Research Institute (CEPRI) of State Grid Corporation of China (SGCC).

\bibliographystyle{IEEEtran}
\bibliography{IEEE_gc_2014}


\end{document}